\documentclass[sigconf]{acmart}

\usepackage{booktabs}

\usepackage{amssymb}
\usepackage{tikz}
\usepackage{pgfplots}
\pgfplotsset{compat=1.18}
\usepgfplotslibrary{fillbetween}
\usepackage{placeins}
\usepackage{float}
\usepackage{tabularx}
\usetikzlibrary{arrows.meta, positioning, fit, backgrounds, calc, shapes.geometric}
\usepackage{enumitem}

\renewcommand{\textsc}[1]{\text{\scshape #1}}
\DeclareMathOperator*{\argmax}{arg\,max}

\setcopyright{none}

\newcommand{\zs}[1]{{\color{black}#1}}

\begin{document}

\title{SkillDroid: Compile Once, Reuse Forever}

\author{Qijia Chen}
\orcid{0000-0003-1038-5461}
\affiliation{%
\department{Department of Computer Science}
  \institution{University of Helsinki}
  \city{Helsinki}
   \country{Finland}
}
\email{qijia.chen@helsinki.fi}

\author{Andrea Bellucci}
\orcid{0000-0003-4035-5271}
\affiliation{%
\department{Department of Computer Science and Engineering}
  \institution{Universidad Carlos III de Madrid}
  \city{Leganés}
  \country{Spain}
  }
\email{abellucc@inf.uc3m.es}

\author{Zhida Sun}
\orcid{0000-0003-4689-986X}
\affiliation{%
\department{CSSE}
  \institution{Shenzhen University}
  \city{Shenzhen}
   \country{China}
}
\email{zhida.sun@szu.edu.cn}

\author{Giulio Jacucci}
\orcid{0000-0002-9185-7928}
\affiliation{
\department{Department of Computer Science}
  \institution{University of Helsinki}
  \city{Helsinki}
  \country{Finland}
  }
\email{giulio.jacucci@helsinki.fi }

\begin{abstract}
LLM-based mobile GUI agents treat every task invocation as an independent reasoning episode, requiring a full LLM inference call at each action step. This per-step dependence makes them stateless: a task completed successfully yesterday is re-derived from scratch today, with no improvement in reliability or speed. We present SkillDroid, a three-layer skill agent that compiles successful LLM-guided GUI trajectories into parameterized skill templates (sequences of UI actions with weighted element locators and typed parameter slots) and replays them on future invocations without any LLM calls. A matching cascade (regex patterns, embedding similarity, and app filtering) routes incoming instructions to stored skills, while a failure-learning layer triggers recompilation when skill reliability degrades. Over a 150-round longitudinal evaluation with systematic instruction variation and controlled perturbations, SkillDroid achieves an 85.3\% success rate (23 percentage points above a stateless LLM baseline) while using 49\% fewer LLM calls. The skill replay mechanism achieves a perfect 100\% success rate across 79 replay rounds at $2.4\times$ the speed of full LLM execution. Most critically, the system improves with use: its success rate converges upward from 87\% to 91\%, while the baseline degrades from 80\% to 44\%.
\end{abstract}

\ccsdesc[500]{Human-centered computing~Ubiquitous and mobile computing systems and tools}
\ccsdesc[500]{Human-centered computing~Interactive systems and tools}
\ccsdesc[500]{Computing methodologies~Artificial intelligence}

\keywords{mobile GUI automation, skill compilation, LLM agents, speculative replay, task automation}

\maketitle

\section{Introduction}

LLM-based GUI agents forget everything between tasks. A user who delegates ``set my morning alarm'' watches the agent reason through the Clock app, locate the right buttons, and succeed. When they repeat the request the next day, the agent starts from scratch: the same app, the same navigation, the same multi-step reasoning, the same latency, 
the same risk of failure. 
Despite having completed the task yesterday, the agent behaves like a first-time user every time.

This statelessness is not a bug in any particular system; it is an architectural property shared by all current LLM-based mobile agents~\cite{wen2023droidbotgpt, wen2024autodroid, zhang2025appagent, wang2024mobileagent, wang2024mobileagentv2}.
Every action step of every task requires a full LLM inference call, making each execution an independent reasoning episode with no accumulated knowledge. 
A recent study on OSWorld found that LLM calls account for 75--94\% of total agent execution time~\cite{abhyankar2025osworld_human}. But the cost is not only computational. 
In our controlled evaluation, a stateless LLM agent's success rate degraded from 80\% to 44\% over 150 rounds as instruction variants grew more diverse, because each round is an independent trial that cannot benefit from past successes or learn from past failures. In other words, stateless agents exhibit non-convergent reliability under repeated use. For end users, this means that repeated delegation does not build trust, because the agent is no more reliable on the hundredth invocation than on the first.

Yet repeated mobile tasks are not independent reasoning problems. ``Set an alarm for 7:30~AM'' and ``Wake me up at 6 tomorrow'' differ in \zs{natural language} phrasing and parameters, but share the same structural pattern: open the Clock app, navigate to Alarms, tap the add button, configure the time, and confirm. This structural regularity 
suggests a fundamentally different approach: rather than reasoning at every step, compile a successful execution into a reusable artifact and replay it mechanically on future invocations.

We present \textbf{SkillDroid}, a three-layer skill agent that shifts mobile GUI automation from opaque, per-step deliberation to predictable, skill-level invocation. When encountering a task for the first time, the system executes it via step-by-step LLM guidance and \emph{compiles} the successful trajectory into a parameterized skill template, a sequence of UI actions with weighted element locators and typed parameter slots (Layer~1). On subsequent invocations, a matching cascade (regex patterns, embedding-based semantic similarity, and target app filtering) identifies the relevant skill and replays its action skeleton directly via the device's accessibility interface, targeting zero LLM calls (Layer~2).
\zs{To handle the dynamic nature of mobile UIs, the system falls back to single-step LLM assistance when deviations occur, and triggers recompilation when a skill's reliability degrades}
(Layer~3). Unlike prior approaches that cache workflows as natural language hints~\cite{wang2024awm}, require LLM calls at every replay step~\cite{lee2024mobilegpt}, or still depend on a lightweight LLM for plan adaptation~\cite{zhang2025apc}, SkillDroid's compiled skills are fully executable without any LLM involvement.

Over a 150-round evaluation spanning 15 task types across diverse Android applications, SkillDroid achieves an 85.3\% success rate, 23 percentage points above a stateless LLM baseline, while using 49\% fewer LLM calls. The skill replay mechanism has a perfect 100\% success rate across 79 rounds at $2.4\times$ the speed of full LLM execution. Critically, the system \zs{demonstrates interactive stability as}
its success rate converges upward from 87\% to 91\% across experimental phases, while the baseline degrades from 80\% to 44\%.

\section{Related Work}

\subsection{LLM-Based Mobile GUI Agents}

Recent work has demonstrated that large language models can act as autonomous controllers for mobile devices by reasoning over UI state at each action step. DroidBot-GPT~\cite{wen2023droidbotgpt} was among the first to query an LLM at every step for Android task automation. AutoDroid~\cite{wen2024autodroid} improves on this by pre-exploring apps to build UI Transition Graphs that are injected as domain-specific memory, achieving 71.3\% task completion on 158 tasks. AppAgent~\cite{zhang2025appagent} builds a knowledge base of UI element descriptions during an exploration phase, retrieving relevant entries at each step to guide GPT-4V's decisions. Mobile-Agent~\cite{wang2024mobileagent} takes a purely vision-based approach using OCR and icon detection on screenshots, while its successor Mobile-Agent-v2~\cite{wang2024mobileagentv2} introduces multi-agent collaboration with separate planning, decision, and reflection modules. On the desktop side, UFO~\cite{zhang2024ufo} coordinates per-application agents for Windows, and OS-Copilot~\cite{wu2024oscopilot} accumulates tools through self-directed learning on general computer tasks. CogAgent~\cite{hong2024cogagent} trains a dedicated 18B-parameter visual language model for high-resolution GUI understanding.

A common architectural property across all these systems is that \emph{every action step of every task invocation requires a full LLM inference call}. This per-step dependence creates two practical problems: (1)~latency is dominated by LLM API round-trips; a recent study on OSWorld found that LLM calls account for 75--94\% of total agent execution time~\cite{abhyankar2025osworld_human}, and (2)~cost scales linearly with the number of task executions, making repeated invocations of the same task unnecessarily expensive. Our work addresses this by \emph{compiling} the LLM's reasoning into a reusable artifact, so that subsequent invocations of the same task type bypass the LLM entirely.

\subsection{Experience Reuse in Simulated and API-Driven Environments}

Prior experience-reuse methods have largely been developed for simulated or API-driven environments, where actions are deterministic and directly executable. These approaches fall into three broad categories:

\emph{(i)~Executable skill libraries in simulated environments.} Voyager~\cite{wang2024voyager} builds an ever-growing library of verified JavaScript functions for Minecraft, indexed by embedding similarity. JARVIS-1~\cite{wang2023jarvis1} augments planning with multimodal memory of past trajectories, GITM~\cite{zhu2023gitm} decomposes goals using text-based knowledge, and Cradle~\cite{tan2024cradle} extends skill curation to desktop software. All operate in environments where UI layouts are fixed and actions always produce the same effect, assumptions that do not hold for real mobile GUIs with their layout variability, dynamic content, and system-level interrupts such as permission dialogs.

\emph{(ii)~Natural language workflow and experience memory.} Rather than storing executable artifacts, several systems cache knowledge as text. Reflexion~\cite{shinn2023reflexion} maintains verbal self-reflections, ExpeL~\cite{zhao2024expel} extracts insights from trajectory comparisons, and Agent Workflow Memory~\cite{wang2024awm} induces reusable workflow routines injected as in-context hints. Agentic Plan Caching~\cite{zhang2025apc} caches structured plan templates for generic agents. In all cases, the LLM still reasons at every step during execution; the cached knowledge serves as guidance, not as a replacement for inference.

\emph{(iii)~Tool and function generation.} LATM~\cite{cai2023latm}, CREATOR~\cite{qian2023creator}, and ToolLLM~\cite{qin2023toolllm} have LLMs generate reusable Python functions or REST API wrappers. These create computational tools, not GUI interaction sequences.

The closest work to ours is MobileGPT~\cite{lee2024mobilegpt}, which caches sub-task procedures in an app memory graph and uses a lighter LLM for slot-filling on recalled sub-tasks. This reduces latency and cost, but each replay step still requires an LLM call for sub-task selection and parameter adaptation. Unlike all of the above, SkillDroid compiles GUI trajectories into \emph{fully executable interaction programs}: parameterized action sequences with weighted element locators, requiring zero LLM calls at replay time.

\section{Method}

\subsection{System Overview}

\textbf{SkillDroid} is a three-layer skill agent that compiles LLM-guided mobile GUI trajectories into reusable parameterized skills. Inspired by the skill library concept in Voyager~\cite{wang2024voyager} and the trajectory-to-program paradigm of \zs{Programming by Demonstration}~\cite{lieberman2001pbd, cypher1993watch}, our core insight is that many mobile tasks share structural patterns: opening an app, navigating to a screen, filling in fields, and confirming. Once the system completes a task via step-by-step LLM guidance, it extracts a parameterized \emph{skill template}, a sequence of UI actions with typed slots, that can be replayed on future invocations without any LLM calls.

The system operates on top of DroidRun\footnote{\url{https://github.com/droidrun/droidrun}}, an open-source Android automation framework that provides accessibility tree capture and ADB-based action execution. Given a natural language instruction (e.g., ``Search for weather in Chrome''), the agent proceeds through three layers:

\begin{itemize}[leftmargin=*]
    \item \textbf{Layer~1 (Compilation):} An LLM executes the task step-by-step, each step consisting of UI state capture, LLM reasoning, action execution, and trajectory recording. On success, a \emph{skill compiler} extracts parameter slots, generates weighted element locators, and stores the resulting skill template in a local SQLite database.

    \item \textbf{Layer~2 (Speculative Replay):} When a new instruction matches an existing skill (via regex pattern matching, embedding-based semantic matching~\cite{reimers2019sentence}, or both), the system replays the skill's action skeleton directly via ADB,\footnote{\url{https://developer.android.com/tools/adb}} targeting zero LLM calls. At each replay step, a \emph{state verifier} checks for UI deviations and an \emph{element finder} locates the target element using weighted feature scoring. If deviations exceed a threshold, the system gracefully falls back to Layer~1.

    \item \textbf{Layer~3 (Failure Learning):} When Layer~2 replay fails, the system records the failure context (step index, deviation type, UI state). If a skill accumulates failures exceeding a 50\% threshold, it is flagged for recompilation on the next matching invocation.
\end{itemize}

\begin{figure*}[t]
    \centering
    \includegraphics[width=0.7\linewidth]{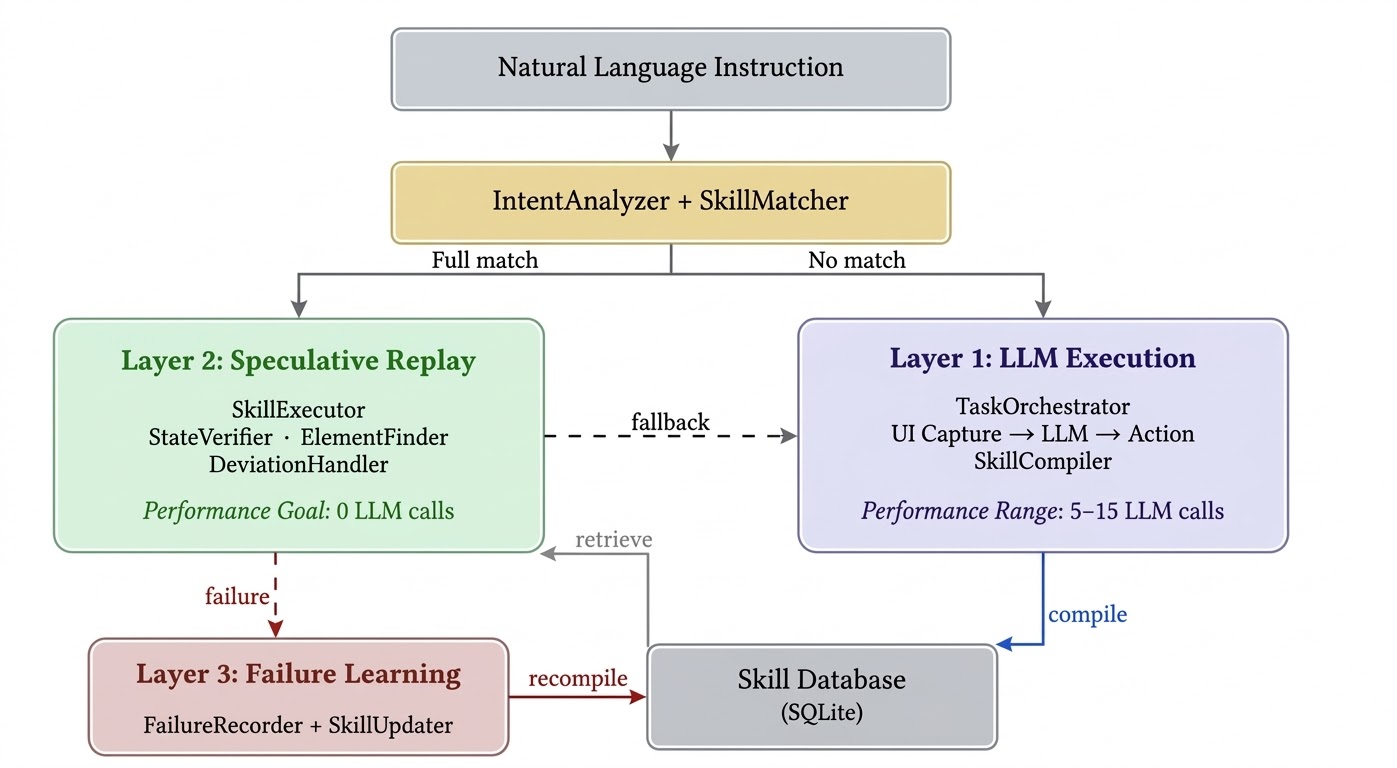}
    \caption{SkillDroid architecture. On a \textsc{Full} match, Layer~2 replays the compiled skeleton (0 LLM calls). On no match, Layer~1 executes via LLM and compiles the trajectory. On replay failure, Layer~1 recovers (dashed) and Layer~3 triggers recompilation when reliability degrades.}
\label{fig:architecture}
\end{figure*}

The \emph{AgentController} orchestrates the three layers: it first queries the \emph{SkillMatcher} for an existing skill; on a full match, it delegates to the \emph{SkillExecutor} for speculative replay; on no match, it invokes the \emph{TaskOrchestrator} for fresh LLM execution, then passes the successful trajectory to the \emph{SkillCompiler}. Figure~\ref{fig:architecture} illustrates the overall architecture.

\subsection{Layer~1: LLM-Guided Execution and Skill Compilation}

A key challenge in LLM-driven GUI automation is that the LLM is an unreliable narrator of its own progress: it may signal task completion while the device state contradicts it, or enter navigation loops without recognizing the impasse. Layer~1 addresses this by wrapping LLM execution in a \emph{verification feedback loop}: a checker-in-the-loop design that converts an open-ended LLM conversation into a bounded, verifiable pipeline whose successful output becomes a reusable skill template.

\subsubsection{Step-by-Step Execution}

The \emph{TaskOrchestrator} drives a loop of up to $N_{\max} = 20$ steps. At each step $t$:

\begin{enumerate}[leftmargin=*]
    \item \textbf{UI State Capture.} The system calls DroidRun's accessibility service to obtain the device's UI hierarchy via the Android Accessibility Framework\footnote{\url{https://developer.android.com/reference/android/accessibilityservice/AccessibilityService}}. We parse the raw accessibility tree (not DroidRun's default formatted text, which strips \texttt{contentDescription} and parent hierarchy) via breadth-first traversal, producing a flat list of UI nodes $\mathcal{U}_t = \{u_1, u_2, \dots, u_n\}$, where each node $u_i$ carries metadata: \texttt{resourceId}, \texttt{text}, \texttt{content\-Description}, \texttt{className}, \texttt{bounds\-InScreen}, and parent/sibling context.

    \item \textbf{LLM Reasoning.} The current UI tree $\mathcal{U}_t$, task instruction $q$, and action history $\mathcal{H}_{<t}$ are formatted into a prompt and sent to the LLM. The LLM returns a structured response $r_t = (a_t, i_t, p_t, \theta_t)$ specifying an action type $a_t \in \{\textsc{Tap}, \textsc{Input}, \textsc{Scroll},\\\textsc{Back}, \textsc{Launch}, \textsc{Done}, \textsc{Fail}\}$, a target element index $i_t$, action parameters $p_t$, and a reasoning trace $\theta_t$.

    \item \textbf{Action Execution.} The selected action is executed via ADB. We tap the center of the target element's \texttt{boundsInScreen} rectangle rather than using DroidRun's index-based tapping, which proved unreliable on emulators.

    \item \textbf{Recording.} Each step is recorded as a trajectory step $s_t = (a_t, \phi_t, \mathcal{U}_t, \theta_t, \text{success}_t)$, where $\phi_t$ denotes the target element's features.
\end{enumerate}

The system includes several safeguards: an \emph{app guardrail} that overrides LLM actions when the foreground application deviates from the target (with a death-loop limit of $G_{\max} = 3$ consecutive overrides); \emph{stale scroll detection} that flags unchanging screens after repeated scrolls; and \emph{mid-execution verification checkpoints} that terminate the loop early when the task is already complete (checked every 3 steps starting from step~5).

\subsubsection{Checker-in-the-Loop Feedback}

When the LLM signals task completion (\textsc{Done} action), the system runs a ground-truth checker $\mathcal{V}$ before accepting. If the checker returns \textsc{Not\_Satisfied}, a feedback message is injected into the LLM conversation:

\begin{quote}
    \emph{``Verification FAILED: \{message\}. The task is NOT complete. Look at the current UI and try again.''}
\end{quote}

\noindent giving the LLM up to $k_{\text{retry}} = 2$ additional attempts. This mechanism catches premature \textsc{Done} signals where the LLM believes the task is complete but the device state does not confirm it.

\subsubsection{Skill Compilation}

Upon successful task completion, the \emph{SkillCompiler} processes the recorded trajectory $\tau = (s_1, s_2, \dots, s_T)$ through three stages:

\paragraph{Slot Extraction.}
An LLM call analyzes the trajectory to identify parameterizable values (task-specific arguments such as contact names, alarm times, or search queries). Each slot $v_j$ is typed ($v_j.\text{type} \in \{\texttt{text}, \texttt{time}, \texttt{phone}, \texttt{url}\}$) and mapped to a placeholder. The instruction is generalized into an intent pattern $\mathcal{P}$; for example, ``Search for weather in Chrome'' becomes ``Search for \texttt{\{search\_query\}} in Chrome''.

\paragraph{Locator Generation.}
For each trajectory step $s_t$ that targets a UI element, the compiler generates a weighted \emph{ElementLocator} $\ell_t$ from the element's features. Drawing on multi-attribute element identification techniques from web testing~\cite{leotta2014reducing, kirinuki2019color}, a locator encodes up to six feature signals from the Android accessibility hierarchy, each weighted by its expected stability across sessions: \texttt{resourceId} ($w{=}0.40$), the developer-assigned identifier that persists across app launches; \texttt{text} ($w{=}0.20$) and \texttt{contentDesc} ($w{=}0.15$), which carry semantic information but vary with dynamic content and i18n; \texttt{className} ($w{=}0.10$), parent context ($w{=}0.10$), and sibling position ($w{=}0.05$), which provide structural disambiguation when primary identifiers are absent. Weights sum to 1.0 and are fixed across all experiments.

Let $F(\ell) \subseteq \{f_1, \dots, f_6\}$ denote the set of features for which locator $\ell$ has a non-null recorded value (the \emph{active features}). At replay time, given a candidate UI element $e$, the element finder computes a match score:

\begin{equation}
    \text{score}(e, \ell) = \frac{\displaystyle\sum_{f \in F(\ell)} w_f \cdot \mathbb{1}\bigl[\textsc{Match}(e.f,\; \ell.f)\bigr]}{\displaystyle\sum_{f \in F(\ell)} w_f}
    \label{eq:locator-score}
\end{equation}

\noindent where $\mathbb{1}[\textsc{Match}(e.f, \ell.f)]$ equals~1 if the element's feature $f$ matches the locator's recorded value. The matching function is feature-dependent: exact string equality for \texttt{resource\-Id} and \texttt{class\-Name}, substring containment for \texttt{text} and \texttt{content\-Desc}, and structural comparison for parent and sibling features. The best-scoring element is selected if it exceeds a threshold:

\begin{equation}
    e^* = \argmax_{e \in \mathcal{U}_t} \text{score}(e, \ell_t), \quad \text{s.t.}\; \text{score}(e^*, \ell_t) \geq \tau
\end{equation}

\noindent where $\tau = \tau_{\text{strict}} = 0.5$ under normal conditions, and $\tau = \tau_{\text{relaxed}} = 0.3$ under minor UI deviations. The strict threshold requires at least the primary identifier (\texttt{resourceId}) plus one secondary feature to match; the relaxed threshold permits matching on secondary features alone when the primary identifier is absent or changed, accepting a controlled risk of misidentification in exchange for replay continuity.

\paragraph{State Descriptor Generation.}
For each step, the compiler captures a \emph{UIStateDescriptor} $\delta_t$, a fingerprint of the expected UI context (current activity, key visible elements), used by Layer~2 to detect deviations during replay.

The compiled \emph{SkillTemplate} $\mathcal{S} = (\mathcal{P}, \{v_j\}, \{(\ell_t, \delta_t, a_t, p_t)\}_{t=1}^{T})$ is persisted to SQLite for future matching.

\subsection{Layer~2: Skill Matching and Speculative Replay}

Simply replaying a recorded macro would fail on the first unexpected popup, layout shift, or dynamic element change, because mobile UIs are inherently non-deterministic across sessions. Layer~2 addresses this with a \emph{graduated deviation response}: at each replay step, a state verifier classifies the mismatch severity, and the system applies progressively stronger interventions: from relaxed element matching, through auto-dismissal of common dialogs, to single-step LLM assistance and full Layer~1 fallback. This design preserves the speed of mechanical replay on the common case while maintaining robustness against the long tail of UI variability.

\subsubsection{Intent Analysis and Skill Matching}

When a new instruction $q'$ arrives, the \emph{IntentAnalyzer} extracts the target application through a multi-pass cascade of explicit context matching (e.g., ``in Chrome''), domain name detection (e.g., \texttt{.com} suffixes), and keyword lookup with longest-match disambiguation (Appendix~\ref{sec:intent-details}). The \emph{SkillMatcher} then attempts to match $q'$ against stored skills $\{\mathcal{S}_1, \dots, \mathcal{S}_K\}$ using a cascade of three strategies:

\paragraph{Regex Pattern Matching.}
Each skill's intent pattern $\mathcal{P}_k$ (e.g., ``Search for \texttt{\{search\_\allowbreak query\}} in Chrome'') is converted to a regex by splitting on slot placeholders and joining with \texttt{(.+)} capture groups. A match directly extracts slot values from the instruction.

\paragraph{Embedding-Based Semantic Matching.}
If regex matching fails, the instruction is encoded using a sentence-transformer model~\cite{reimers2019sentence} (all-MiniLM-L6-v2, 22M parameters) and compared against pre-computed skill embeddings via cosine similarity. Each skill's intent pattern is embedded with slot placeholders stripped. Skills scoring above a threshold of $\tau_{\text{sem}} = 0.40$ are considered candidates. This threshold is intentionally permissive to maximize recall over structurally diverse paraphrases (e.g., ``Wake me up at 6'' vs.\ ``Set an alarm for \{\}''): false positives are caught by a subsequent LLM confirmation call that verifies the match and extracts slot values, costing one additional API invocation but preventing incorrect skill replay.

\paragraph{Target App Filtering.}
Candidates are filtered to ensure the skill's target application matches the instruction's resolved target app.
Match results are classified as \textsc{Full} (complete match with extracted slots), \textsc{Partial} (app matches but instruction diverges), or \textsc{No\_Match}.

\subsubsection{Speculative Replay}

On a \textsc{Full} match with skill $\mathcal{S}_k$ and extracted slot values $\{v_j \mapsto c_j\}$, the \emph{SkillExecutor} replays the skill's step sequence $(\ell_1, \delta_1, a_1, p_1), \dots, (\ell_T, \delta_T, a_T, p_T)$. At each step $t$:

\paragraph{State Verification.}
The \emph{StateVerifier} compares the current UI state against the expected descriptor $\delta_t$. Deviations are classified into four severity levels $d \in \{\textsc{None}, \textsc{Minor}, \textsc{Moderate}, \textsc{Major}\}$:

\begin{itemize}[leftmargin=*]
    \item \textsc{None}: All checks pass; continue with $\tau = \tau_{\text{strict}}$.
    \item \textsc{Minor}: Same app, some elements shifted; use $\tau = \tau_{\text{relaxed}}$.
    \item \textsc{Moderate}: Same app, unexpected dialog (e.g., permission prompt); the \emph{DeviationHandler} attempts auto-dismissal by scanning for dismiss-like buttons (``Allow'', ``OK'', ``Skip'', etc.) using word-boundary regex matching (\verb|\b...\b|) to avoid false positives (e.g., ``ok'' in ``booking'').
    \item \textsc{Major}: Different app entirely; abort replay and fall back to Layer~1.
\end{itemize}

\paragraph{Element Finding.}
The \emph{ElementFinder} scores each element $e \in \mathcal{U}_t$ against locator $\ell_t$ using Equation~\ref{eq:locator-score}. For slot-bound steps (e.g., a text input field whose locator records \texttt{\{search\_query\}}), the actual slot value $c_j$ is substituted before matching.

\paragraph{Step-Level LLM Fallback.}
If element finding fails at step $t$ (e.g., a dynamic time picker whose elements differ across invocations), a single LLM call is made to resolve that step, after which replay resumes. This is bounded by $B_{\text{consec}} = 2$ consecutive and $B_{\text{total}} = 5$ total step-level fallbacks per execution; exceeding either triggers a full Layer~1 fallback.

\paragraph{Step Skipping.}
If the current UI state already reflects step $t$'s expected post-condition (e.g., a permission dialog from the original trajectory is absent), step $t$ is skipped. This handles trajectory noise from permission grants and retries during initial compilation.

\subsubsection{Fallback and Recompilation}

When Layer~2 replay fails, the system falls back to Layer~1 with a \emph{PriorContext} containing the completed speculative steps, so the LLM does not repeat already-completed actions. If the Layer~1 fallback succeeds, the recovery trajectory is compiled as a \emph{new} skill rather than overwriting the original, since the original skill may still be valid for other instruction variants that do not trigger the same deviation. The original skill's failure count is incremented, and recompilation is triggered only when its failure rate exceeds the threshold (Section~\ref{sec:layer3}).

\subsection{Layer~3: Failure Learning}
\label{sec:layer3}

Initial skill compilations are often noisy: they may contain steps for permission dialogs that only appear once, retry sequences from transient failures, or suboptimal navigation paths. Rather than treating the first successful trajectory as ground truth, Layer~3 monitors skill reliability over time and progressively replaces underperforming skills with cleaner versions compiled from later, more representative trajectories.

The \emph{FailureRecorder} tracks each Layer~2 failure with its context: the step index at which replay diverged, the deviation severity, the UI state at failure, and whether Layer~1 recovery succeeded. The \emph{SkillUpdater} analyzes failure patterns to generate \emph{GuardConditions}, pre-checks that can be evaluated before replay to predict likely failures.

Importantly, skills flagged with \texttt{needs\_recompile} are \emph{not} filtered out during matching: the \emph{SkillMatcher} still returns them as candidates, and the \emph{AgentController} decides whether to attempt replay or force fresh execution. This prevents the system from losing track of partially useful skills that may succeed under certain instruction variants.

A skill's failure rate $r_{\text{fail}} = n_{\text{fail}} / (n_{\text{succ}} + n_{\text{fail}})$ is tracked continuously. Skills with $r_{\text{fail}} > 0.5$ are flagged for recompilation; on the next matching invocation, the system forces a fresh Layer~1 execution to produce an updated skill template. Recompilation is capped at $V_{\max} = 3$ versions to avoid futile loops on inherently unreliable tasks.

\section{Experimental Setup}

\subsection{Tasks, Instructions, and Perturbations}

The evaluation comprises 150 rounds executed on 15 task types covering diverse GUI interaction patterns: multi-field forms (Contacts, Calendar), time pickers (Clock), address bar navigation (Chrome), toggle switches, sliders, and deep Settings navigation, text editing (Keep Notes), and app launching (Appendix~\ref{app:details}, Table~\ref{tab:tasks}). Tasks range from 1-step toggles (T9: WiFi, T10: Airplane) to 8+ step multi-field forms (T1: Create Contact, T8: Calendar Event), with 0 to 4 parameter slots.

Each task type has instructions at four variation levels: \emph{Compile}~(C), the canonical phrasing used for initial skill compilation (e.g., ``Set an alarm for 7:30 AM''); \emph{Low}~(L), same pattern with new parameters (``Set an alarm for 9:00 AM''); \emph{Medium}~(M), paraphrased (``Create an alarm for 6:15 AM''); and \emph{High}~(H), colloquial (``Wake me up at 6 tomorrow''). This progression tests the matching cascade from exact regex to embedding-based semantic matching.

To test robustness, Phase~4 introduces controlled perturbations applied before task execution (Table~\ref{tab:perturbations}).

\begin{table}[t]
\centering
\small
\begin{tabular}{lll}
\toprule
\textbf{Perturbation} & \textbf{Method} & \textbf{Expected Impact} \\
\midrule
App chooser dialog & Inject intent chooser & Moderate deviation \\
Clear app data     & \texttt{pm clear} target app & Welcome flow triggers \\
Revoke permission  & \texttt{pm revoke} permission & Permission dialog \\
\bottomrule
\end{tabular}
\caption{Perturbations applied during Phase~4 robustness testing.}
\label{tab:perturbations}
\vspace{-30px}
\end{table}

\subsection{Evaluation Protocol}

\subsubsection{Main Experiment (150 Rounds)}

The experiment consists of 150 rounds organized into 5 phases using gpt-4o-mini as the LLM (Appendix~\ref{app:details}, Table~\ref{tab:phases}): P1 (R1--15) compiles the initial skill library using canonical instructions; P2 (R16--45) tests Layer~2 replay with low-variation instructions; P3 (R46--75) introduces paraphrased and colloquial variants to test semantic matching; P4 (R76--105) adds controlled perturbations to test recovery; and P5 (R106--150) measures steady-state convergence with mixed variation levels.

Each round follows the sequence: device reset (force-stop target app, return to home screen, wait for UI stabilization) $\rightarrow$ optional setup (e.g., pre-create a contact for deletion tasks) $\rightarrow$ optional perturbation $\rightarrow$ task execution $\rightarrow$ ground-truth verification. The emulator is restarted every 30~rounds to prevent accessibility service degradation.

\subsubsection{Baseline Experiment (150 Rounds)}

As a controlled comparison, we run the identical 150 tasks using only the \emph{TaskOrchestrator} (Layer~1): the same LLM, prompts, action execution, reset procedures, and ground-truth verification, but with no skill matching, no speculative replay, and no skill compilation. This isolates the contribution of the skill system from the underlying LLM capability.

\subsubsection{Supplementary Experiment}

To evaluate generalization beyond the main configuration, we conduct a 75-round \textbf{cross-model} experiment using gpt-4o, covering Phases~1--3 (compilation, exact reuse, and semantic reuse). These phases exercise the model-dependent aspects of the system (compilation quality, slot extraction accuracy, and LLM reasoning during Layer~1 execution), while Phases~4--5 primarily test model-independent mechanisms (perturbation recovery, failure learning) already evaluated in the main experiment.

\subsection{Ground-Truth Verification}

We do not rely on the LLM's self-reported task completion. Each task type has a dedicated programmatic checker $\mathcal{V}_k$ that queries device state via ADB shell commands (e.g., querying \texttt{content://contacts}, reading \texttt{settings get global}, parsing UI trees) to verify the expected outcome. A round is scored as successful only when the checker returns \textsc{Verified}. The full set of verification methods is listed in Appendix~\ref{app:verify}, Table~\ref{tab:checkers}.

\subsection{Metrics}

We report success rate (fraction of rounds passing ground-truth verification), LLM calls per round, end-to-end latency, Layer~2/Layer~1 speedup ratio, skill match rate, 0-LLM rate (rounds with zero LLM execution calls), and fallback rate (Layer~2 attempts that degrade to Layer~1). Formal definitions are provided in Appendix~\ref{sec:metrics-definitions}.

\subsection{Implementation}

The system is implemented in Python~3.12 atop DroidRun~v0.5. The skill database uses synchronous SQLite. Semantic matching uses sentence-transformers~\cite{reimers2019sentence} with the all-MiniLM-L6-v2 model~\cite{wang2020minilm} (22M parameters, 80\,MB). All experiments run on a Windows~11 host driving an Android emulator (Pixel~9a, API~35, DroidRun Portal~v0.5.6) via ADB, with coordinate-based tapping at ${\sim}100$\,ms per action. The LLM is accessed through the OpenAI API\footnote{\url{https://platform.openai.com/docs/models}} (gpt-4o-mini for the main experiment, gpt-4o for the supplementary) with temperature $T = 0.2$ and a per-call timeout of 60\,s. Each round is limited to $N_{\max} = 20$ steps, and the emulator is restarted every 30 rounds to prevent accessibility service degradation. We note that ADB-based action execution (${\sim}100$\,ms per tap) is a deliberate portability trade-off: a native Android \texttt{Accessibility\-Service} integration using \texttt{performAction()} can reduce per-action latency to ${\sim}4$\,ms, further improving Layer~2 speedup at the cost of requiring on-device APK installation. We discuss this trade-off in Section~\ref{sec:limitations}.
To support reproducibility, we will release code and evaluation scripts upon acceptance.
\section{Results}

\subsection{Skill Replay Effectiveness}

\begin{table}[b]
\centering
\small
\begin{tabular}{lrrrr}
\toprule
\textbf{Execution Path} & $n$ & \textbf{Success} & $\bar{C}_{\text{LLM}}$ & $\bar{L}$ (s) \\
\midrule
L2: Pure replay          & 35  & 100\%  &   0.0 & 36.0 \\
L2 + semantic match      & 32  & 100\%  &   1.0 & 54.7 \\
L2 + step-level fallback & 12  & 100\%  &   5.0 & 50.9 \\
\midrule
\emph{All Layer~2 variants} & \emph{79} & \emph{100\%} & \emph{1.2} & \emph{45.1} \\
\midrule
L2 $\rightarrow$ L1 fallback & 29  & 75.9\% & 10.1 & 113.2 \\
L1: Fresh LLM            & 42  & 64.3\% &  11.6 & 84.0 \\
\midrule
\textbf{All rounds}      & 150 & 85.3\% &   5.8 & 69.6 \\
\bottomrule
\end{tabular}
\caption{Performance by execution layer. When Layer~2 replay succeeds without full fallback (top three rows), the success rate is 100\% across 79 rounds, with a mean of only 1.2 LLM calls and $2.4\times$ speedup over the baseline.}
\label{tab:layer-dist}
\end{table}

\begin{table}[t]
\centering
\small
\begin{tabular}{lcc}
\toprule
\textbf{Metric} & \textbf{SkillDroid} & \textbf{Baseline} \\
\midrule
Success rate              & 128/150 (85.3\%)  & 93/150 (62.0\%) \\
Mean LLM calls / round   & 5.8               & 11.3 \\
Mean latency / round      & 69.0\,s           & 84.1\,s \\
Pure L2 rounds (0 LLM)   & 35 (23.3\%)       & --- \\
\bottomrule
\end{tabular}
\caption{Aggregate comparison (150 rounds, gpt-4o-mini).}
\label{tab:main-results}
\end{table}

The central claim of SkillDroid is that compiled skills can be reliably replayed without LLM involvement. Table~\ref{tab:layer-dist} validates this claim by breaking down performance across execution layers.

When a skill matches and Layer~2 replay completes without requiring a full Layer~1 fallback, the success rate is \textbf{100\% across all 79 rounds} (52.7\% of the experiment). Of these, 35 rounds use pure skeleton replay with \textbf{zero LLM calls} and a mean latency of 35.4\,s, a $2.4\times$ speedup compared to the baseline's overall mean latency of 84.1\,s (Table~\ref{tab:main-results}). Another 32 rounds require a single LLM call for semantic match confirmation, and 12 rounds invoke limited step-level LLM fallbacks for dynamic UI elements (e.g., time pickers) while still completing via replay.

The skill matching cascade is effective: among the 135 post-compilation rounds (P2--P5), 107 (79.3\%) receive a \textsc{Full} match. Regex pattern matching handles same-structure instructions (Low variation), while embedding-based semantic matching successfully captures paraphrased and colloquial instructions (Medium and High variations). The three-strategy cascade (regex $\rightarrow$ embedding $\rightarrow$ app filter) ensures broad coverage without sacrificing precision.

The L2 $\rightarrow$ L1 fallback path (29 rounds, 19.3\%) has a 75.9\% success rate, indicating that the fallback mechanism recovers the majority of replay failures. Its high latency (113.2\,s) reflects the cost of an attempted replay followed by a full LLM execution; however, the recovery trajectory is compiled as a new skill, preventing the same failure from recurring.

Table~\ref{tab:scope-analysis} isolates the skill framework's intrinsic reliability by progressively removing LLM capability bottlenecks. SkillDroid's 22 failures are dominated by two task types where the LLM itself cannot complete execution (T1: 0\% for both systems; T11: deep-nested navigation). Excluding these, SkillDroid achieves 97.6\%. Excluding by failure mode (removing only rounds where the LLM exhausted all 20 steps), only 2 of 130 non-timeout rounds fail, yielding 98.5\%. The skill replay mechanism itself has \emph{never} failed across 79 rounds.

\begin{table}[b]
\centering
\small
\begin{tabular}{lc}
\toprule
\textbf{Scope} & \textbf{Success Rate} \\
\midrule
All 150 rounds                    & 128/150 (85.3\%) \\
Excl.\ T1 \& T11 (LLM bottlenecks) & 124/127 (97.6\%) \\
Excl.\ step-limit exhaustion     & 128/130 (98.5\%) \\
\midrule
Layer~2 replay (no full fallback) & \phantom{0}79/79\phantom{0} (100\%) \\
\bottomrule
\end{tabular}
\caption{SkillDroid success rate by analysis scope. Progressively excluding LLM capability bottlenecks reveals that the skill framework itself achieves 98.5\% reliability; Layer~2 replay has never failed.}
\label{tab:scope-analysis}
\end{table}

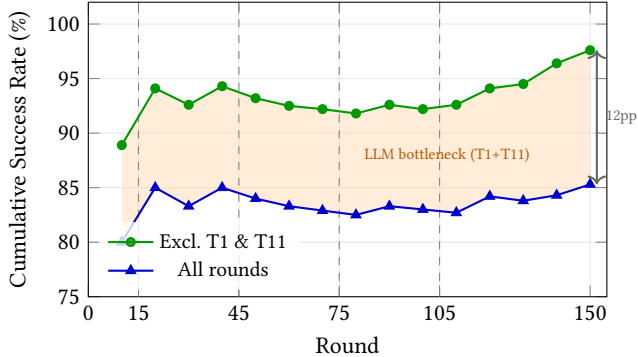
\begin{figure}[t]
\centering
\begin{tikzpicture}
\begin{axis}[
    width=\columnwidth,
    height=5.5cm,
    xlabel={Round},
    ylabel={Cumulative Success Rate (\%)},
    xmin=0, xmax=155,
    ymin=75, ymax=102,
    xtick={0,15,45,75,105,150},
    ytick={75,80,85,90,95,100},
    legend style={at={(0.02,0.02)}, anchor=south west, font=\small, draw=none, fill=white, fill opacity=0.8, text opacity=1},
    grid=major,
    grid style={gray!20},
    every axis plot/.append style={thick},
    clip=false,
]
\draw[gray, dashed, thin] (axis cs:15,75) -- (axis cs:15,102);
\draw[gray, dashed, thin] (axis cs:45,75) -- (axis cs:45,102);
\draw[gray, dashed, thin] (axis cs:75,75) -- (axis cs:75,102);
\draw[gray, dashed, thin] (axis cs:105,75) -- (axis cs:105,102);
\addplot[green!60!black, mark=*, mark size=1.5pt, name path=A] coordinates {
    (10,88.9) (20,94.1) (30,92.6) (40,94.3) (50,93.2) (60,92.5)
    (70,92.2) (80,91.8) (90,92.6) (100,92.2) (110,92.6) (120,94.1)
    (130,94.5) (140,96.4) (150,97.6)
};
\addplot[blue!80!black, mark=triangle*, mark size=1.8pt, name path=B] coordinates {
    (10,80.0) (20,85.0) (30,83.3) (40,85.0) (50,84.0) (60,83.3)
    (70,82.9) (80,82.5) (90,83.3) (100,83.0) (110,82.7) (120,84.2)
    (130,83.8) (140,84.3) (150,85.3)
};
\addplot[orange!30, opacity=0.5] fill between[of=A and B];
\draw[<->, thick, black!60] (axis cs:152,97.6) -- node[right, font=\scriptsize] {12pp} (axis cs:152,85.3);
\node[right, font=\scriptsize, orange!70!black] at (axis cs:80,88) {LLM bottleneck (T1+T11)};
\legend{Excl.\ T1 \& T11, All rounds}
\end{axis}
\end{tikzpicture}
\caption{Intrinsic reliability of the skill framework. Excluding the two task types where the LLM itself cannot complete execution (T1, T11), SkillDroid's cumulative success rate reaches 97.6\%. The shaded gap quantifies the cost of LLM capability limitations on the overall metric.}
\label{fig:intrinsic-reliability}
\end{figure}

Figure~\ref{fig:intrinsic-reliability} visualizes this decomposition as cumulative success curves. The green curve (excluding T1 \& T11) rises to 97.6\% and remains nearly flat throughout, confirming that the skill framework itself is highly reliable once LLM capability bottlenecks are removed. The gap between the two curves, consistently ${\sim}12$ percentage points, quantifies the exact cost of LLM limitations on the overall metric.

Table~\ref{tab:variation} breaks down performance by instruction variation level. SkillDroid maintains stable success rates across all levels (80--89\%), while the baseline degrades from 80.0\% (Compile) to 52.2\% (High). The matching cascade distributes load appropriately: Low-variation rounds are predominantly handled by pure L2 regex replay (20 of 43), Medium and High variations trigger embedding-based semantic matching (16 and 15 of their respective totals), and unmatched instructions fall through to Layer~1. Across all post-compilation variation levels, the L2 replay rate remains consistent (57--60\%), confirming that the matching cascade generalizes effectively to paraphrased and colloquial instructions.

\begin{table}[t]
\centering
\small
\begin{tabular}{lrcccc}
\toprule
\textbf{Variation} & $n$ & \multicolumn{2}{c}{\textbf{Success Rate}} & \multicolumn{2}{c}{$\bar{C}_{\text{LLM}}$} \\
\cmidrule(lr){3-4} \cmidrule(lr){5-6}
                   &     & SkillDroid & Baseline & SkillDroid & Baseline \\
\midrule
Compile (P1)       & 15  & 86.7\% & 80.0\% &  9.2 &  9.4 \\
Low                & 43  & 86.0\% & 62.8\% &  4.7 & 10.6 \\
Medium             & 46  & 80.4\% & 65.2\% &  6.6 & 11.1 \\
High               & 46  & 89.1\% & 52.2\% &  5.0 & 13.0 \\
\bottomrule
\end{tabular}
\caption{Performance by instruction variation level. SkillDroid remains stable (80--89\%) regardless of instruction difficulty, while the baseline degrades significantly for harder variations. SkillDroid's LLM usage drops sharply after compilation, confirming skill reuse across all variation levels.}
\label{tab:variation}
\end{table}

\FloatBarrier
\subsection{Learning Dynamics}

The phase-by-phase comparison (Table~\ref{tab:phase-trends}) reveals a striking divergence. SkillDroid's success rate is stable across phases and \emph{improves} in P5 (91.1\%), reflecting skill library maturation: accumulated skills cover more instruction variants, and recompiled skills replace noisy initial compilations. Its LLM usage drops from P1 (9.2 calls, all Layer~1) to P2 (6.5 calls) and continues declining to P5 (4.2 calls), confirming that skill reuse progressively displaces LLM computation.

The baseline degrades monotonically from 80.0\% to 44.4\%. Viewed through 30-round sliding windows, the decline is stark: 76.7\% (R1--30) $\rightarrow$ 80.0\% (R31--60) $\rightarrow$ 60.0\% (R61--90) $\rightarrow$ 50.0\% (R91--120) $\rightarrow$ 43.3\% (R121--150). Without a skill library, every round is an independent LLM execution with no memory, so the same errors recur, and the later phases' harder instruction variants (paraphrased and colloquial phrasings, perturbations) expose the LLM's brittleness. The baseline's LLM call count remains constant at ${\sim}11$ across all phases, confirming the absence of any learning effect.

This divergence is the central empirical finding: the skill compilation mechanism transforms a stateless LLM executor into a system that improves with use. By P5, the gap reaches 47 percentage points, with SkillDroid completing most tasks via efficient skill replay while the baseline continues to fail on the same tasks it handled earlier.

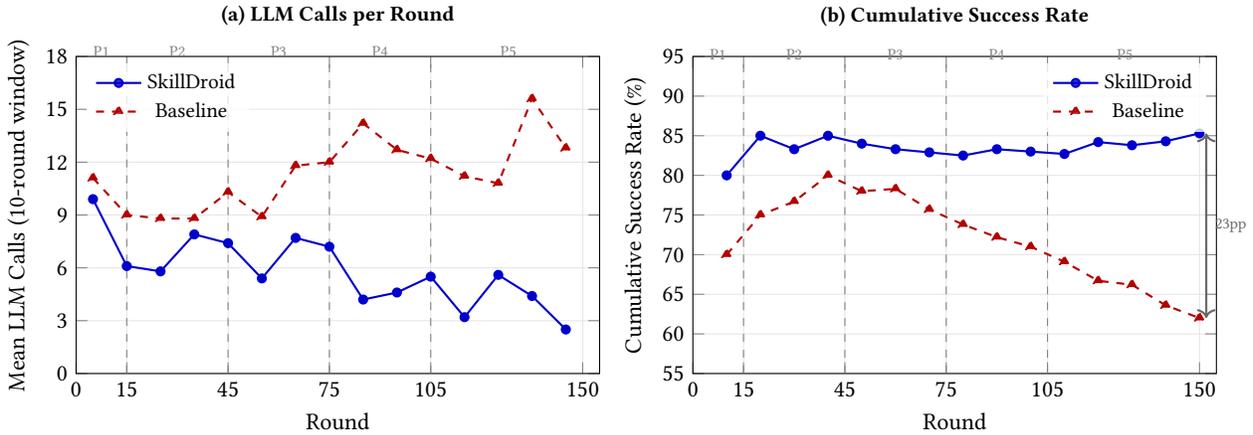
\begin{figure*}[t]
\centering
\begin{tikzpicture}
\begin{scope}[xshift=0cm]
\begin{axis}[
    width=0.48\textwidth,
    height=5.8cm,
    xlabel={Round},
    ylabel={Mean LLM Calls (10-round window)},
    xmin=0, xmax=155,
    ymin=0, ymax=18,
    xtick={0,15,45,75,105,150},
    ytick={0,3,6,9,12,15,18},
    legend style={at={(0.02,0.98)}, anchor=north west, font=\small, draw=none, fill=white, fill opacity=0.8, text opacity=1},
    grid=major,
    grid style={gray!20},
    every axis plot/.append style={thick},
    title={\textbf{(a) LLM Calls per Round}},
    title style={at={(0.5,1.02)}, font=\small},
    clip=false,
]
\draw[gray, dashed, thin] (axis cs:15,0) -- (axis cs:15,18);
\draw[gray, dashed, thin] (axis cs:45,0) -- (axis cs:45,18);
\draw[gray, dashed, thin] (axis cs:75,0) -- (axis cs:75,18);
\draw[gray, dashed, thin] (axis cs:105,0) -- (axis cs:105,18);
\node[above, font=\scriptsize, gray] at (axis cs:7.5,17.5) {P1};
\node[above, font=\scriptsize, gray] at (axis cs:30,17.5) {P2};
\node[above, font=\scriptsize, gray] at (axis cs:60,17.5) {P3};
\node[above, font=\scriptsize, gray] at (axis cs:90,17.5) {P4};
\node[above, font=\scriptsize, gray] at (axis cs:128,17.5) {P5};
\addplot[blue!80!black, mark=*, mark size=1.5pt] coordinates {
    (5,9.9) (15,6.1) (25,5.8) (35,7.9) (45,7.4) (55,5.4)
    (65,7.7) (75,7.2) (85,4.2) (95,4.6) (105,5.5) (115,3.2)
    (125,5.6) (135,4.4) (145,2.5)
};
\addplot[red!70!black, mark=triangle*, mark size=1.8pt, dashed] coordinates {
    (5,11.1) (15,9.0) (25,8.8) (35,8.8) (45,10.3) (55,8.9)
    (65,11.8) (75,12.0) (85,14.2) (95,12.7) (105,12.2) (115,11.2)
    (125,10.8) (135,15.6) (145,12.8)
};
\legend{SkillDroid, Baseline}
\end{axis}
\end{scope}
\begin{scope}[xshift=8.2cm]
\begin{axis}[
    width=0.48\textwidth,
    height=5.8cm,
    xlabel={Round},
    ylabel={Cumulative Success Rate (\%)},
    xmin=0, xmax=155,
    ymin=55, ymax=95,
    xtick={0,15,45,75,105,150},
    ytick={55,60,65,70,75,80,85,90,95},
    legend style={at={(0.98,0.98)}, anchor=north east, font=\small, draw=none, fill=white, fill opacity=0.8, text opacity=1},
    grid=major,
    grid style={gray!20},
    every axis plot/.append style={thick},
    title={\textbf{(b) Cumulative Success Rate}},
    title style={at={(0.5,1.02)}, font=\small},
    clip=false,
]
\draw[gray, dashed, thin] (axis cs:15,55) -- (axis cs:15,95);
\draw[gray, dashed, thin] (axis cs:45,55) -- (axis cs:45,95);
\draw[gray, dashed, thin] (axis cs:75,55) -- (axis cs:75,95);
\draw[gray, dashed, thin] (axis cs:105,55) -- (axis cs:105,95);
\node[above, font=\scriptsize, gray] at (axis cs:7.5,93.5) {P1};
\node[above, font=\scriptsize, gray] at (axis cs:30,93.5) {P2};
\node[above, font=\scriptsize, gray] at (axis cs:60,93.5) {P3};
\node[above, font=\scriptsize, gray] at (axis cs:90,93.5) {P4};
\node[above, font=\scriptsize, gray] at (axis cs:128,93.5) {P5};
\addplot[blue!80!black, mark=*, mark size=1.5pt] coordinates {
    (10,80.0) (20,85.0) (30,83.3) (40,85.0) (50,84.0) (60,83.3)
    (70,82.9) (80,82.5) (90,83.3) (100,83.0) (110,82.7) (120,84.2)
    (130,83.8) (140,84.3) (150,85.3)
};
\addplot[red!70!black, mark=triangle*, mark size=1.8pt, dashed] coordinates {
    (10,70.0) (20,75.0) (30,76.7) (40,80.0) (50,78.0) (60,78.3)
    (70,75.7) (80,73.8) (90,72.2) (100,71.0) (110,69.1) (120,66.7)
    (130,66.2) (140,63.6) (150,62.0)
};
\draw[<->, thick, black!60] (axis cs:152,85.3) -- node[right, font=\scriptsize] {23pp} (axis cs:152,62.0);
\legend{SkillDroid, Baseline}
\end{axis}
\end{scope}
\end{tikzpicture}
\caption{Learning curves over 150 rounds. \textbf{(a)}~10-round rolling average of LLM calls: SkillDroid's cost drops from ${\sim}10$ (P1) to ${\sim}3$ (P5) as the skill library grows, while the baseline stays flat at ${\sim}11$--13 and rises for harder instruction variants. \textbf{(b)}~Cumulative success rate: SkillDroid holds steady at ${\sim}84\%$; the baseline degrades from 77\% to 62\%. Dashed lines mark phase boundaries. The widening scissors pattern is visual evidence that skill compilation transforms a stateless executor into a system that improves with use.}
\label{fig:learning-curves}
\end{figure*}

Figure~\ref{fig:learning-curves} visualizes this divergence. The left panel shows LLM calls per round (10-round rolling average): SkillDroid's curve declines from ${\sim}10$ in P1 to ${\sim}3$ in P5 as the skill library grows, while the baseline remains flat at ${\sim}11$--13 and even increases as later phases introduce harder instruction variants. The right panel shows cumulative success rate: SkillDroid holds steady at ${\sim}84\%$ throughout, while the baseline degrades continuously from ${\sim}77\%$ to 62\%, with the gap widening at every phase boundary. Together, these curves demonstrate that SkillDroid \emph{learns from experience}: each compiled skill reduces future LLM reliance and stabilizes long-run reliability.

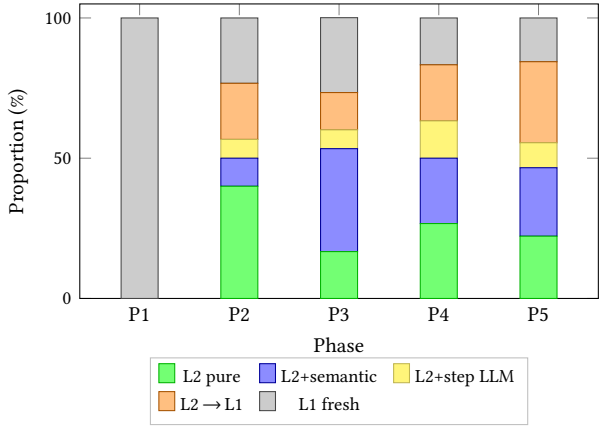
\begin{figure}[t]
\centering
\begin{tikzpicture}
\begin{axis}[
    ybar stacked,
    width=\columnwidth,
    height=5.5cm,
    ylabel={Proportion (\%)},
    ylabel style={font=\small},
    xlabel={Phase},
    xlabel style={font=\small},
    ymin=0, ymax=105,
    xtick=data,
    xticklabels={P1, P2, P3, P4, P5},
    xticklabel style={font=\small},
    yticklabel style={font=\footnotesize},
    bar width=14pt,
    enlarge x limits=0.15,
    legend style={
        at={(0.5,-0.20)}, anchor=north,
        legend columns=3,
        font=\footnotesize,
        draw=gray!50,
        /tikz/every even column/.append style={column sep=4pt},
    },
]
\addplot[fill=green!55, draw=green!70!black] coordinates
  {(1,0) (2,40.0) (3,16.7) (4,26.7) (5,22.2)};
\addlegendentry{L2 pure}
\addplot[fill=blue!45, draw=blue!60!black] coordinates
  {(1,0) (2,10.0) (3,36.7) (4,23.3) (5,24.4)};
\addlegendentry{L2+semantic}
\addplot[fill=yellow!65, draw=yellow!70!black] coordinates
  {(1,0) (2,6.7) (3,6.7) (4,13.3) (5,8.9)};
\addlegendentry{L2+step LLM}
\addplot[fill=orange!50, draw=orange!65!black] coordinates
  {(1,0) (2,20.0) (3,13.3) (4,20.0) (5,28.9)};
\addlegendentry{L2\,$\rightarrow$\,L1}
\addplot[fill=gray!40, draw=gray!55!black] coordinates
  {(1,100) (2,23.3) (3,26.7) (4,16.7) (5,15.6)};
\addlegendentry{L1 fresh}
\end{axis}
\end{tikzpicture}
\caption{Execution layer distribution by phase. P1 is entirely Layer~1 (skill compilation). Over subsequent phases, Layer~2 variants (green, blue, yellow) progressively displace Layer~1 (gray), reflecting skill library growth. By P5, 84\% of rounds use some form of skill replay.}
\label{fig:layer-evolution}
\end{figure}

Figure~\ref{fig:layer-evolution} reveals the mechanism behind this convergence: the execution layer mix shifts dramatically across phases. In P1, 100\% of rounds use Layer~1 (fresh LLM execution). By P5, Layer~1 drops to 16\% while Layer~2 variants collectively account for 84\% of rounds, with pure replay (22\%), semantic matching (24\%), and step-level LLM fallback (9\%) each contributing. The growing L2$\to$L1 fallback proportion in P4--P5 (20--29\%) reflects increased instruction diversity rather than skill degradation: harder paraphrases occasionally trigger deviations, but the fallback recovers 76\% of these cases.

\subsection{Overall Comparison}

Tables~\ref{tab:main-results} and~\ref{tab:phase-trends} summarize the aggregate and phase-by-phase comparisons.

\begin{table}[b]
\centering
\small
\begin{tabular}{lcccc}
\toprule
\textbf{Phase} & \multicolumn{2}{c}{\textbf{SkillDroid}} & \multicolumn{2}{c}{\textbf{Baseline}} \\
\cmidrule(lr){2-3} \cmidrule(lr){4-5}
               & Success & $\bar{C}_{\text{LLM}}$ & Success & $\bar{C}_{\text{LLM}}$ \\
\midrule
P1: Compilation     & 86.7\% &   9.2 & 80.0\% &  9.4 \\
P2: Exact Reuse     & 83.3\% &   6.5 & 80.0\% &  9.4 \\
P3: Semantic Reuse  & 83.3\% &   6.5 & 66.7\% & 11.0 \\
P4: Robustness      & 80.0\% &   5.2 & 56.7\% & 12.9 \\
P5: Steady State    & 91.1\% &   4.2 & 44.4\% & 12.6 \\
\bottomrule
\end{tabular}
\caption{Phase-by-phase success rate and mean LLM calls. SkillDroid converges upward (86.7\% $\rightarrow$ 91.1\%) while the baseline degrades monotonically (80.0\% $\rightarrow$ 44.4\%).}
\label{tab:phase-trends}
\end{table}


SkillDroid outperforms the baseline by 23 percentage points in success rate while using 49\% fewer LLM calls and completing tasks 18\% faster on average. The largest gains appear in tasks with complex but structurally repeatable UI flows: T2 (Delete Contact, +62pp), T13 (Font Size, +57pp), T6 (Chrome Search, +46pp), and T12 (DND Toggle, +44pp). These tasks involve multi-step navigation that the LLM frequently mishandles within 20 steps, but a compiled skill replays reliably. Tasks where both systems achieve 100\% (T5, T8, T10) represent simple, well-structured flows where the LLM already performs well, yet SkillDroid still provides efficiency gains (63--94\% fewer LLM calls) without sacrificing reliability. The full per-task breakdown is provided in Appendix, Table~\ref{tab:per-task}.

\subsection{Supplementary Experiments}

\subsubsection{Cross-Model: gpt-4o}

To test whether skill reuse benefits generalize across LLM capabilities, we run Phases~1--3 (75 rounds) of the same task sequence using gpt-4o (Table~\ref{tab:supplementary}).

\begin{table}[t]
\centering
\small
\begin{tabular}{lccc}
\toprule
\textbf{Configuration} & \textbf{Success} & $\bar{C}_{\text{LLM}}$ & $\bar{L}$ (s) \\
\midrule
SkillDroid (gpt-4o-mini, 150R)  & 85.3\% & 5.8 & 69.0 \\
Baseline (gpt-4o-mini, 150R)    & 62.0\% & 11.3 & 84.1 \\
\midrule
SkillDroid (gpt-4o, 75R)        & 90.7\% & 4.9 & 44.5 \\
\bottomrule
\end{tabular}
\caption{Supplementary experiment. gpt-4o over Phases~1--3 (75 rounds) achieves higher compilation reliability (93\% in P1) and comparable skill reuse benefits.}
\label{tab:supplementary}
\end{table}

With gpt-4o, SkillDroid achieves 90.7\% success across 75 rounds (P1--P3). The stronger base model improves Layer~1 compilation reliability (93\% in P1 vs.\ 87\% for gpt-4o-mini), producing cleaner skill templates. LLM calls drop from 6.6 in P1 to 4.2 in P2 as skills are reused, and pure Layer~2 replay latency averages 29.0\,s. Both P2 and P3 sustain 90\% success, confirming that the skill reuse mechanism provides consistent benefits across model tiers: a more capable LLM compiles better skills, which then replay just as efficiently.

\subsection{Failure Analysis}
\label{sec:failure-analysis}

\subsubsection{SkillDroid Failures}

Of SkillDroid's 22 failures, 19 (86\%) are concentrated in two task types: T1 (Create Contact, 0/13) and T11 (Toggle Location, 4/10). Both systems struggle with these tasks: T1 fails at 0\% for the baseline as well, due to the Contacts app's highly variable multi-field form, and T11's deeply nested Settings menus pose challenges even with skill assistance (though SkillDroid improves over the baseline's 10\% to 40\%).

On the remaining 13 task types, SkillDroid achieves \textbf{124/127 (97.6\%)} success. When T1 and T11 are excluded (Appendix~\ref{sec:appendix-excl}), the baseline's success rate drops to 67.6\% while SkillDroid maintains 88.2\%, suggesting a 21-point gap driven by the skill compilation mechanism.

The 3 remaining failures (one each in T6, T7, T14) occur during L2$\rightarrow$L1 fallback recovery. T6 was triggered by the \texttt{pm clear} perturbation, which erased Chrome's data and produced a first-run welcome flow that consumed the 20-step budget during Layer~1 recovery. T7 failed without perturbation when the LLM typed an incorrect URL during fallback and exhausted the step budget attempting to correct. T14 failed because the target app became temporarily unreachable during the experiment session. The same skill templates succeed on other rounds (T6: 12/13, T7: 9/10, T14: 8/9).

\subsubsection{Baseline Failures}

The baseline's 57 failures follow a different pattern. The dominant failure mode is exceeding the maximum step limit: 46 of 57 failures (80.7\%) occur because the LLM enters navigation loops (repeatedly tapping incorrect elements or scrolling without progress) until the 20-step budget is exhausted. Another 8 failures (14.0\%) involve the LLM reporting task completion while the ground-truth checker rejects the result. The remaining 3 failures stem from LLM output parsing errors (2) and a transient device connectivity issue (1). Unlike SkillDroid, the baseline has no mechanism to learn from or avoid these repeated failures.

\subsubsection{Perturbation Robustness}

Phase~4 includes 10 rounds with controlled perturbations. App chooser dialog injection (4 rounds) was handled at 100\% by the DeviationHandler's auto-dismiss mechanism with zero LLM calls, validating the \textsc{Moderate} deviation handling design. The \texttt{pm clear} perturbation (4 rounds) succeeded in 3 cases via Layer~1 fallback, failing only when Chrome's first-run welcome flow consumed too many steps. Permission revocation (2 rounds) coincided with the persistently failing Contacts tasks, preventing isolation. Overall, 7/10 perturbation rounds succeeded, demonstrating meaningful robustness against unexpected UI state changes.

\subsubsection{Skill Recompilation (Layer~3)}

Over 150 rounds, the skill database accumulated 25 skill entries across 15 task types, of which 10 (40\%) are recompilations (version $> 1$) and 5 reached the maximum $V_{\max} = 3$. Three skills (Set Alarm, v1: 40\% success; Note Creation, v1: 33\%; Timer Setting) were flagged with \texttt{needs\_recompile} during the experiment. All three stabilized at 100\% success after recompilation, as cleaner trajectories (free of permission dialogs and retry artifacts) replaced the noisy originals. This progressive refinement contributes to the upward convergence observed in P5.

\subsection{Component Analysis}
\label{sec:component-analysis}

To validate which design decisions contribute to SkillDroid's improvement over the baseline, we analyze three key mechanisms in detail (Appendix~\ref{sec:component-details}). The weighted element locator achieves a 91.1\% effective hit rate across 325 element-find attempts, with \texttt{resourceId} (weight 0.40, present in 72.8\% of locators) serving as the dominant matching signal. Step-level resilience mechanisms (step skipping and single-step LLM fallback) rescued 26 rounds (17.3\%) from full Layer~1 fallback, each saving an estimated 60\,s. The checker-in-the-loop design eliminates false-completion failures entirely: 14\% of baseline failures stem from the LLM declaring ``Done'' prematurely, while SkillDroid experiences zero such failures because Layer~2 replay completes on a mechanical condition, not an LLM judgment.

\section{Discussion}
\label{sec:discussion}

Beyond aggregate accuracy, the \zs{primary} HCI implication of SkillDroid is that compiling successful trajectories changes repeated-task delegation.
\zs{It shifts the interaction model} from one-shot, \zs{opaque} reasoning to execution over reusable skills. This shift matters not just for system performance, but also for user experience \zs{of AI agents}. Our results suggest benefits for predictability, latency on repeated tasks, user understanding, and repairability. We focus on three implications that are specific to compiled-skill agents.

\paragraph{Predictability and trust calibration.}
The strongest user-facing consequence of skill compilation is not simply higher average success, but more predictable repeated execution. Layer~2 replay succeeds in 79/79 rounds, and the framework reaches 98.5\% reliability once failures caused by upstream LLM capability limits are excluded (Table~\ref{tab:layer-dist}). Over time, SkillDroid's overall success rate converges upward from 87\% to 91\%, while the stateless baseline degrades from 80\% to 44\% (Table~\ref{tab:phase-trends}). 
For repeated tasks, this means a successful interaction is not a fresh reasoning episode each time. It is a reusable execution path that can become more dependable with use. We do not claim to measure user trust directly, but these results suggest a stronger basis for calibrated trust \cite{lee2004trust,parasuraman1997humans} that stateless agents cannot offer, where each success is an independent event with no guarantee of future stability.

\paragraph{Provenance, control, and privacy.}
Compiled-skill agents also create a new transparency requirement: users need to know how the system completed a task, not just whether it succeeded. The same instruction may be completed through pure replay (35.4\,s, 0 LLM calls), semantic-matched replay (54.7\,s, 1 call), step-level fallback (50.9\,s, 5 calls), or fresh Layer~1 execution. These paths differ meaningfully in speed, reliability, and degree of open-ended reasoning (Table~\ref{tab:layer-dist}). 
This suggests that interfaces for compiled-skill agents should expose execution provenance.
\zs{This suggests that interfaces for compiled-skill agents should expose execution provenance, for example, by indicating whether the system is replaying a known skill, adapting one, or reasoning from scratch.}
Following Horvitz's principle of mixed-initiative interaction~\cite{horvitz1999principles}, silent execution may be appropriate for pure replay, while step-level adaptation or full fallback may warrant user awareness or confirmation for sensitive tasks.
Provenance also determines privacy exposure. A persistent concern with LLM-based GUI agents is that every action step transmits UI content to external APIs.  Pure layer 2 sends no user data to external LLM APIs. It keeps instructions and UI state entirely on-device.

\paragraph{Repairability and inspectability.}
Skill compilation makes repeated-task automation more repairable than opaque one-shot execution. Step skipping rescued 14 rounds by skipping 22 stale steps, all successfully, showing that replay can tolerate benign UI drift without full failure. When skills did become noisy, recompilation repaired them: three skills that initially achieved only 33--50\% success stabilized at 100\% after cleaner trajectories were recompiled. The checker-in-the-loop design also removes an entire failure category: 14\% of baseline failures come from false LLM completion signals, whereas SkillDroid has none because replay completion is mechanical rather than judgment-based (Appendix~\ref{sec:component-details}).
These properties align with a long-standing principle in HCI: that automated behaviors should be inspectable and correctable by end users~\cite{lieberman2001pbd,cypher1993watch}. Compiled skills, with their parameterized slots and explicit step sequences, are closer to this ideal than opaque LLM reasoning traces.


\section{Limitations and Future Work}
\label{sec:limitations}

We view reusable agent skills as a promising substrate for future delegation interfaces. Compiling trajectories into persistent skills shifts repeated delegation from one-off prompting toward accumulated capability. This raises open design questions: how should users inspect skills, repair incorrect matches, manage skill libraries, set confirmation policies, and understand replay outcomes? These interaction challenges are important directions for future work. Beyond them, the current system has several technical limitations.

\paragraph{Task Coverage.}
Our evaluation prioritizes depth over breadth: 150 rounds with systematic instruction variation (four levels per task) and controlled perturbations, rather than a larger task catalogue with single-shot execution. This design is necessary to evaluate skill compilation, matching robustness, and longitudinal convergence, properties invisible in one-off benchmarks. Extending to broader task coverage (e.g., multi-app workflows, drag-and-drop interactions) and physical devices with varying screen sizes is a natural next step.

\paragraph{Complex Form Limitations.}
The T1 (Create Contact) failure at 0\% for both systems exposes a fundamental limitation: highly dynamic multi-field forms with expandable sections, inconsistent element ordering, and context-dependent ``Add field'' controls cannot be reliably handled by either LLM-guided execution or skill replay within a 20-step budget. Addressing such tasks may require hierarchical skill decomposition (compiling sub-skills for individual form sections) or vision-based UI understanding that captures spatial layout beyond the accessibility tree.

\paragraph{Text-Only UI Perception.}
The current system relies exclusively on accessibility tree text. It cannot interpret visual cues such as icon semantics, color-coded states, or spatial grouping that a human user would leverage. Integrating screenshot-based perception could improve both Layer~1 compilation quality and Layer~2 state verification accuracy, particularly for apps with poor accessibility annotations.

\paragraph{ADB Latency Overhead.}
ADB-based action execution (${\sim}100$\,ms per tap) is approximately $25\times$ slower than native on-device accessibility APIs (${\sim}4$\,ms), limiting the practical speedup of skill replay. A native integration would significantly reduce Layer~2 latency, but at the cost of requiring app installation and system permissions on the target device. Our ADB-based approach trades action speed for deployment flexibility, requiring only a standard USB/TCP connection.

\paragraph{Single-Language and Single-LLM.}
All instructions are in English, and the main experiment uses a single LLM family (OpenAI gpt-4o-mini/gpt-4o). While the 75-round gpt-4o supplementary experiment confirms cross-model generalizability, testing with open-source models and multilingual instructions remains future work.

\paragraph{Skill Composition.}
Each skill targets a single task within one application. Real-world automation often requires multi-step workflows spanning multiple apps (e.g., ``Find a restaurant on Maps and share the address via WhatsApp''). Extending the framework to support skill composition, chaining atomic skills into compound workflows with inter-skill data passing, is a natural and promising direction.

\section{Conclusion}

We presented SkillDroid, a three-layer skill agent that compiles LLM-guided mobile GUI trajectories into parameterized, LLM-free skill templates for autonomous replay. Over 150 rounds spanning 15 task types, SkillDroid achieves an 85.3\% success rate, 23 percentage points above a stateless LLM baseline, while using 49\% fewer LLM calls. The skill replay mechanism itself has a \emph{perfect} 100\% success rate across 79 rounds, with pure replay completing in zero LLM calls at $2.4\times$ the baseline's speed. Most importantly, the system improves with use: success rate converges upward from 87\% to 91\% while the baseline degrades from 80\% to 44\%, demonstrating that skill compilation transforms a memoryless LLM executor into one that learns from experience.
More broadly, we see reusable agent skills as an interaction substrate for mobile delegation systems. In which users can come to expect not just one-off task completion, but progressively more predictable, reusable, and controllable behavior over tim.

\bibliographystyle{ACM-Reference-Format}
\bibliography{references}

\appendix

\section*{Appendix}

\section{Task and Instruction Details}\label{app:details}

\begin{table}[H]
\centering
\footnotesize
\begin{tabularx}{\columnwidth}{lXll}
\toprule
\textbf{Category} & \textbf{Tasks} & \textbf{App} & \textbf{Pattern} \\
\midrule
Contacts  & T1: Create, T2: Delete & Contacts & Multi-field form \\
Clock     & T3: Timer, T4: Alarm, T5: Del.\ alarm & Clock & Time picker \\
Browser   & T6: Search, T7: Open URL & Chrome & Address bar \\
Calendar  & T8: Create event & Calendar & Multi-field form \\
Settings  & T9--T10: WiFi, Airplane, T11: Location, T12: DND, T13: Font & Settings & Toggle, slider, deep nav. \\
Notes     & T14: Create note & Keep Notes & Text editor \\
Launch    & T15: Open app & Any & App drawer \\
\bottomrule
\end{tabularx}
\caption{The 15 task types used in evaluation. Tasks range from 1-step toggles (T9, T10) to 8+ step multi-field forms (T1, T8), with 0 to 4 parameter slots.}
\label{tab:tasks}
\end{table}

\begin{table}[H]
\centering
\footnotesize
\begin{tabular}{@{}lllp{2.2cm}@{}}
\toprule
\textbf{Phase} & \textbf{Rounds} & \textbf{Variation} & \textbf{Purpose} \\
\midrule
P1: Compilation     & R1--R15    & C          & Build skill library \\
P2: Exact Reuse     & R16--R45   & L          & Test L2 replay \\
P3: Semantic Reuse  & R46--R75   & M, H       & Test semantic matching \\
P4: Robustness      & R76--R105  & Mixed + perturb. & Test recovery \\
P5: Steady State    & R106--R150 & L, M, H    & Measure convergence \\
\bottomrule
\end{tabular}
\caption{Five-phase structure of the 150-round main experiment.}
\label{tab:phases}
\end{table}

\section{Intent Analyzer Details}
\label{sec:intent-details}

When a new instruction $q'$ arrives, the \emph{IntentAnalyzer} extracts the target application through a multi-pass cascade:

\begin{enumerate}[leftmargin=*]
    \item \textbf{Explicit context matching}: Regex patterns detect explicit app references (e.g., ``in Chrome'', ``using Maps'') to extract the target app directly.
    \item \textbf{Domain name detection}: If no explicit context is found, a regex scan for domain suffixes (e.g., \texttt{.com}, \texttt{.org}) infers a browser target, ensuring ``Open youtube.com'' routes to Chrome rather than the YouTube app.
    \item \textbf{Keyword matching}: As a fallback, the instruction is matched against a known app dictionary using longest-keyword-wins to resolve ambiguities (e.g., ``phone number'' matches Contacts, not Dialer).
\end{enumerate}

\section{Ground-Truth Verification Methods}\label{app:verify}

\begin{table}[H]
\centering
\small
\begin{tabular}{ll}
\toprule
\textbf{Task Type} & \textbf{Verification Method} \\
\midrule
Create/Delete Contact & Query \texttt{content://contacts} for target name \\
Set Alarm / Timer     & Parse Clock UI tree for target time \\
Chrome Search / URL   & Verify foreground app + URL bar content \\
WiFi / Airplane / etc.\ & Read \texttt{settings get global/secure} values \\
Calendar Event        & Query \texttt{content://calendar/events} \\
Font Size             & Read \texttt{settings get system font\_scale} \\
Create Note           & Verify app foreground + note title visible \\
Open App              & Check target package in foreground \\
\bottomrule
\end{tabular}
\caption{Ground-truth verification methods. Each checker returns \textsc{Verified}, \textsc{Not\_Satisfied}, or \textsc{Check\_Error}.}
\label{tab:checkers}
\end{table}

\section{Metric Definitions}
\label{sec:metrics-definitions}

\begin{itemize}[leftmargin=*]
    \item \textbf{Success Rate}: $\frac{|\{r : \mathcal{V}(r) = \textsc{Verified}\}|}{|\text{rounds}|}$, the fraction of rounds passing ground-truth verification.

    \item \textbf{LLM Calls} ($C_{\text{LLM}}$): Total LLM API invocations per round. $C_{\text{LLM}} = 0$ for pure Layer~2 replay; $C_{\text{LLM}} = 1$ for semantic match + replay; $C_{\text{LLM}} \in [5, 15]$ for full Layer~1 execution.  

    \item \textbf{End-to-End Latency} ($L$): Wall-clock time from task start to verification completion, in seconds.

    \item \textbf{Speedup}: $\frac{\bar{L}_{\text{L1}}}{\bar{L}_{\text{L2}}}$, the ratio of mean Layer~1 latency to mean Layer~2 latency for the same task type.

    \item \textbf{Skill Match Rate}: $\frac{|\{r : \text{match}(r) = \textsc{Full}\}|}{|\text{post-P1 rounds}|}$, the fraction of post-compilation rounds receiving a full skill match.

    \item \textbf{0-LLM Rate}: $\frac{|\{r : C_{\text{LLM}}^{\text{exec}}(r) = 0\}|}{|\text{rounds}|}$, the fraction of rounds completed with zero LLM execution calls.

    \item \textbf{Fallback Rate}: $\frac{|\{r : \text{L2} \rightarrow \text{L1}\}|}{|\text{L2 attempts}|}$, the fraction of Layer~2 attempts that degrade to full Layer~1 execution.
\end{itemize}

\section{Per-Task Breakdown}

\begin{table}[H]
\centering
\small
\begin{tabular}{lcccc}
\toprule
\textbf{Task} & \multicolumn{2}{c}{\textbf{Success Rate}} & \multicolumn{2}{c}{\textbf{Mean $C_{\text{LLM}}$}} \\
\cmidrule(lr){2-3} \cmidrule(lr){4-5}
               & SkillDroid & Baseline & SkillDroid & Baseline \\
\midrule
T1: Create contact    &  0/13~~(0\%)   &  0/13~~(0\%)   & 19.8 & 19.6 \\
T2: Delete contact    &  8/8~~(100\%)  &  3/8~~(38\%)   &  3.1 & 13.2 \\
T3: Set timer         &  9/9~~(100\%)  &  7/9~~(78\%)   &  3.7 &  6.3 \\
T4: Set alarm         & 12/12~(100\%)  &  9/12~(75\%)   &  8.1 & 11.2 \\
T5: Delete alarm      &  9/9~~(100\%)  &  9/9~~(100\%)  &  4.6 &  7.2 \\
T6: Chrome search     & 12/13~(92\%)   &  6/13~(46\%)   &  4.5 & 14.2 \\
T7: Open URL          &  9/10~(90\%)   &  6/10~(60\%)   &  3.3 & 11.6 \\
T8: Calendar event    & 13/13~(100\%)  & 13/13~(100\%)  &  2.0 &  7.4 \\
T9: WiFi toggle       & 10/10~(100\%)  &  8/10~(80\%)   &  0.5 & 10.9 \\
T10: Airplane toggle  &  9/9~~(100\%)  &  9/9~~(100\%)  &  0.4 &  6.2 \\
T11: Location toggle  &  4/10~(40\%)   &  1/10~(10\%)   & 12.6 & 19.8 \\
T12: DND toggle       &  9/9~~(100\%)  &  5/9~~(56\%)   &  2.9 & 10.7 \\
T13: Font size        &  7/7~~(100\%)  &  3/7~~(43\%)   &  7.9 & 14.6 \\
T14: Create note      &  8/9~~(89\%)   &  7/9~~(78\%)   &  5.9 &  5.9 \\
T15: Open app         &  9/9~~(100\%)  &  7/9~~(78\%)   &  3.8 &  8.2 \\
\bottomrule
\end{tabular}
\caption{Per-task comparison. SkillDroid matches or exceeds the baseline on all 15 tasks. The largest gains appear in tasks with complex but repeatable UI flows (T2: +62pp, T13: +57pp, T6: +46pp, T12: +44pp). Tasks where both systems fail (T1, T11) reflect LLM capability limitations rather than skill system failures.}
\label{tab:per-task}
\end{table}

\section{Supplementary Figures}

\subsection{Fair Comparison Excluding LLM Bottleneck Tasks}
\label{sec:appendix-excl}

\begin{figure*}[t]
\centering
\begin{tikzpicture}
\begin{scope}[xshift=0cm]
\begin{axis}[
    width=0.48\textwidth,
    height=5.8cm,
    xlabel={Round},
    ylabel={Mean LLM Calls (10-round window)},
    xmin=0, xmax=155,
    ymin=0, ymax=20,
    xtick={0,15,45,75,105,150},
    ytick={0,4,8,12,16,20},
    legend style={at={(0.02,0.98)}, anchor=north west, font=\small, draw=none, fill=white, fill opacity=0.8, text opacity=1},
    grid=major,
    grid style={gray!20},
    every axis plot/.append style={thick},
    title={\textbf{(a) LLM Calls per Round}},
    title style={at={(0.5,1.02)}, font=\small},
    clip=false,
]
\draw[gray, dashed, thin] (axis cs:15,0) -- (axis cs:15,20);
\draw[gray, dashed, thin] (axis cs:45,0) -- (axis cs:45,20);
\draw[gray, dashed, thin] (axis cs:75,0) -- (axis cs:75,20);
\draw[gray, dashed, thin] (axis cs:105,0) -- (axis cs:105,20);
\node[above, font=\scriptsize, gray] at (axis cs:7.5,19.5) {P1};
\node[above, font=\scriptsize, gray] at (axis cs:30,19.5) {P2};
\node[above, font=\scriptsize, gray] at (axis cs:60,19.5) {P3};
\node[above, font=\scriptsize, gray] at (axis cs:90,19.5) {P4};
\node[above, font=\scriptsize, gray] at (axis cs:128,19.5) {P5};
\addplot[blue!80!black, mark=*, mark size=1.5pt] coordinates {
    (7,7.1) (19,3.7) (30,6.8) (40,7.0) (55,3.3) (66,8.2)
    (79,5.0) (94,2.0) (106,2.7) (118,5.3) (130,2.8) (141,3.3)
};
\addplot[red!70!black, mark=triangle*, mark size=1.8pt, dashed] coordinates {
    (7,8.3) (19,7.6) (30,8.4) (40,7.9) (53,7.7) (63,10.7)
    (73,13.4) (84,12.8) (96,11.3) (108,10.7) (119,10.1) (130,10.6)
    (140,17.2)
};
\legend{SkillDroid, Baseline}
\end{axis}
\end{scope}
\begin{scope}[xshift=8.2cm]
\begin{axis}[
    width=0.48\textwidth,
    height=5.8cm,
    xlabel={Round},
    ylabel={Cumulative Success Rate (\%)},
    xmin=0, xmax=155,
    ymin=60, ymax=102,
    xtick={0,15,45,75,105,150},
    ytick={60,65,70,75,80,85,90,95,100},
    legend style={at={(0.98,0.98)}, anchor=north east, font=\small, draw=none, fill=white, fill opacity=0.8, text opacity=1},
    grid=major,
    grid style={gray!20},
    every axis plot/.append style={thick},
    title={\textbf{(b) Cumulative Success Rate}},
    title style={at={(0.5,1.02)}, font=\small},
    clip=false,
]
\draw[gray, dashed, thin] (axis cs:15,60) -- (axis cs:15,102);
\draw[gray, dashed, thin] (axis cs:45,60) -- (axis cs:45,102);
\draw[gray, dashed, thin] (axis cs:75,60) -- (axis cs:75,102);
\draw[gray, dashed, thin] (axis cs:105,60) -- (axis cs:105,102);
\node[above, font=\scriptsize, gray] at (axis cs:7.5,100.5) {P1};
\node[above, font=\scriptsize, gray] at (axis cs:30,100.5) {P2};
\node[above, font=\scriptsize, gray] at (axis cs:60,100.5) {P3};
\node[above, font=\scriptsize, gray] at (axis cs:90,100.5) {P4};
\node[above, font=\scriptsize, gray] at (axis cs:128,100.5) {P5};
\addplot[blue!80!black, mark=*, mark size=1.5pt] coordinates {
    (10,100.0) (20,100.0) (30,96.2) (40,94.4) (50,90.9) (60,90.6)
    (70,88.9) (80,87.5) (90,87.7) (100,86.7) (110,87.8) (120,86.9)
    (130,87.1) (140,88.1) (150,88.2)
};
\addplot[red!70!black, mark=triangle*, mark size=1.8pt, dashed] coordinates {
    (10,87.5) (20,88.2) (30,88.5) (40,88.9) (50,88.6) (60,88.7)
    (70,84.1) (80,81.9) (90,80.2) (100,78.9) (110,77.6) (120,74.8)
    (130,73.3) (140,69.8) (150,67.6)
};
\draw[<->, thick, black!60] (axis cs:152,88.2) -- node[right, font=\scriptsize] {21pp} (axis cs:152,67.6);
\legend{SkillDroid, Baseline}
\end{axis}
\end{scope}
\end{tikzpicture}
\caption{Learning curves excluding T1 and T11 (136 rounds). \textbf{(a)}~10-round rolling average of LLM calls: SkillDroid's cost drops from ${\sim}7$ to ${\sim}3$, while the baseline fluctuates at ${\sim}8$--17, with the same flat-to-rising pattern as the full experiment. \textbf{(b)}~Cumulative success rate: both systems start near ${\sim}90\%$ (confirming T1/T11 drove the initial gap), but the baseline degrades to 67.6\% while SkillDroid holds at 88.2\%. The 21pp final gap is entirely attributable to the skill compilation mechanism.}
\label{fig:excl-t1t11-comparison}
\end{figure*}
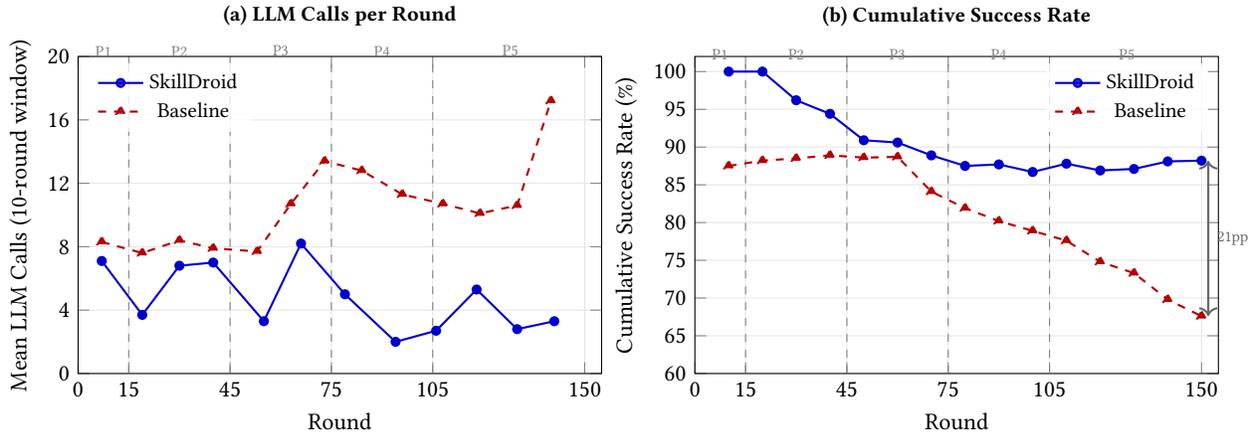

As discussed in Section~\ref{sec:failure-analysis}, T1 (Create Contact) and T11 (Toggle Location) are tasks where the LLM itself cannot reliably complete execution; both systems fail for the same underlying reasons. Figure~\ref{fig:excl-t1t11-comparison} removes these 23 rounds (13 T1 + 10 T11) and compares the two systems on the remaining 136 rounds, mirroring the structure of Figure~\ref{fig:learning-curves}.

With LLM bottleneck tasks removed, the baseline's initial success rate rises to ${\sim}89\%$, comparable to SkillDroid's early performance, confirming that T1 and T11 account for much of the early gap. However, the baseline still degrades monotonically from 89\% to 67.6\%, while SkillDroid stabilizes at 88.2\%. The LLM calls panel shows an even sharper contrast: SkillDroid's cost drops to ${\sim}3$ calls by P5, while the baseline fluctuates at ${\sim}10$--17. The final success rate gap of 21 percentage points demonstrates that the skill compilation mechanism provides substantial value \emph{even on tasks where the LLM is fundamentally capable}.

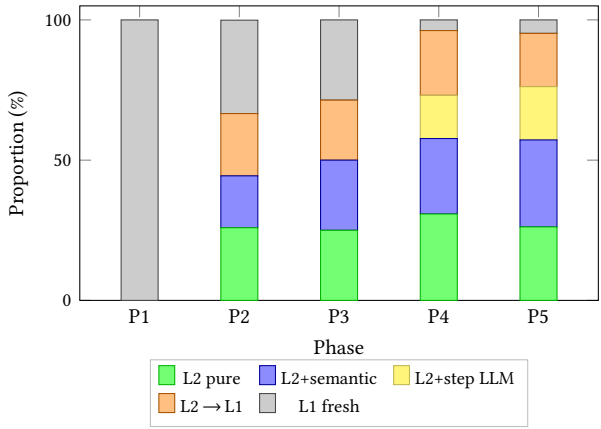
\begin{figure}[H]
\centering
\begin{tikzpicture}
\begin{axis}[
    ybar stacked,
    width=\columnwidth,
    height=5.5cm,
    ylabel={Proportion (\%)},
    ylabel style={font=\small},
    xlabel={Phase},
    xlabel style={font=\small},
    ymin=0, ymax=105,
    xtick=data,
    xticklabels={P1, P2, P3, P4, P5},
    xticklabel style={font=\small},
    yticklabel style={font=\footnotesize},
    bar width=14pt,
    enlarge x limits=0.15,
    legend style={
        at={(0.5,-0.20)}, anchor=north,
        legend columns=3,
        font=\footnotesize,
        draw=gray!50,
        /tikz/every even column/.append style={column sep=4pt},
    },
]
\addplot[fill=green!55, draw=green!70!black] coordinates
  {(1,0) (2,25.9) (3,25.0) (4,30.8) (5,26.2)};
\addlegendentry{L2 pure}
\addplot[fill=blue!45, draw=blue!60!black] coordinates
  {(1,0) (2,18.5) (3,25.0) (4,26.9) (5,31.0)};
\addlegendentry{L2+semantic}
\addplot[fill=yellow!65, draw=yellow!70!black] coordinates
  {(1,0) (2,0.0) (3,0.0) (4,15.4) (5,19.0)};
\addlegendentry{L2+step LLM}
\addplot[fill=orange!50, draw=orange!65!black] coordinates
  {(1,0) (2,22.2) (3,21.4) (4,23.1) (5,19.0)};
\addlegendentry{L2\,$\rightarrow$\,L1}
\addplot[fill=gray!40, draw=gray!55!black] coordinates
  {(1,100) (2,33.3) (3,28.6) (4,3.8) (5,4.8)};
\addlegendentry{L1 fresh}
\end{axis}
\end{tikzpicture}
\caption{Execution layer distribution by phase, excluding T1 and T11. Compared to Figure~\ref{fig:layer-evolution}, Layer~1 fresh execution is dramatically reduced in P4 (13.3\% $\rightarrow$ 3.8\%) and P5 (6.7\% $\rightarrow$ 4.8\%), confirming that on LLM-capable tasks, skill replay achieves near-complete coverage ($>$95\%) by the later phases.}
\label{fig:layer-evolution-excl}
\end{figure}

Figure~\ref{fig:layer-evolution-excl} shows the execution layer distribution by phase after excluding T1 and T11. The most notable change is in P4, where Layer~1 fresh execution drops from 13.3\% (Figure~\ref{fig:layer-evolution}) to just 3.8\%, indicating that the majority of P4's original Layer~1 rounds were T1/T11 attempts. By P4--P5, Layer~2 variants account for 96.2\% and 95.2\% of rounds respectively, confirming near-complete skill coverage on LLM-capable tasks.

\section{Component Analysis}
\label{sec:component-details}

The aggregate results demonstrate that SkillDroid outperforms the baseline, but do not reveal \emph{which design decisions} contribute to this improvement. This section validates three key mechanisms: the weighted element locator, the step-level resilience cascade, and the checker-in-the-loop verification.

\subsection{Weighted Element Locator}

The element locator is the core of Layer~2 replay: each compiled skill step stores a weighted feature vector (Equation~\ref{eq:locator-score}), and at replay time the executor searches the live UI tree for the best-matching element. Across 108 Layer~2 rounds, we observe 382 element-find attempts. Of these, 296 (77.5\%) are resolved by the locator. The remaining 86 divide into 57 stale-step skips (trajectory steps from compilation artifacts like permission dialogs that no longer appear) and 29 genuine failures that trigger fallback. Excluding stale-step skips, the locator's effective hit rate is \textbf{296/325 = 91.1\%}.

Table~\ref{tab:locator-features} shows the feature composition of all 92 compiled locators across 25 skills. The \texttt{resourceId} feature (weight 0.40) is present in 72.8\% of locators and is the dominant matching signal: when present, it alone contributes 47--62\% of the match score, typically sufficient for the relaxed threshold (0.3). The 27.2\% of locators \emph{without} \texttt{resourceId} depend on the lower-weight combination of \texttt{className} + \texttt{parent} + \texttt{siblingIndex} (combined weight 0.25), making them inherently more fragile. Notably, skills composed entirely of \texttt{resourceId}-bearing locators (e.g., Delete Alarm, Calendar Event) achieve near-perfect L2 completion, while skills with fragile locators (e.g., Set Alarm, Font Size) account for most step-level fallbacks.

\begin{table}[H]
\centering
\small
\begin{tabular}{lccl}
\toprule
\textbf{Feature} & \textbf{Weight} & \textbf{Present in} & \textbf{Role} \\
\midrule
\texttt{resourceId}    & 0.40 & 67/92 (72.8\%) & Primary anchor \\
\texttt{text}          & 0.20 & 37/92 (40.2\%) & Button labels, slot values \\
\texttt{contentDesc}   & 0.15 & 35/92 (38.0\%) & Icon semantics \\
\texttt{className}     & 0.10 & 92/92 (100\%)  & Type tiebreaker \\
\texttt{parent}        & 0.10 & 91/92 (98.9\%) & Context disambiguation \\
\texttt{siblingIndex}  & 0.05 & 92/92 (100\%)  & Positional last-resort \\
\bottomrule
\end{tabular}
\caption{Feature presence in 92 compiled element locators. \texttt{resourceId} is the most discriminative signal; locators lacking it (27.2\%) rely on weaker structural features and are more prone to matching failure.}
\label{tab:locator-features}
\end{table}

The two-tier threshold design (strict 0.5, relaxed 0.3) provides an important safety net: the most common locator combination (\texttt{resourceId} + \texttt{text} + structural features, 33.7\% of locators) scores 0.47 on \texttt{resourceId} alone, just below the strict threshold, but passes the relaxed threshold comfortably. This enables successful matching even when slot-substituted text differs from the compiled template.

\subsection{Step-Level Resilience}

Layer~2 replay is designed to degrade gracefully rather than fail catastrophically. Two mechanisms, \emph{step skipping} and \emph{step-level LLM fallback}, allow replay to recover from local mismatches without falling back to a full Layer~1 re-execution.

\paragraph{Step skipping.} When a skill step's locator finds no match in the current UI but a \emph{later} step's locator does match, the executor skips ahead, treating the missing steps as compilation artifacts (e.g., permission dialogs or retries from the original LLM execution that do not appear during replay). Across 150 rounds, 14 rounds used step skipping (22 total steps skipped), all completing successfully. This mechanism is particularly valuable for skills compiled from noisy Layer~1 trajectories before recompilation cleans them.

\paragraph{Step-level LLM fallback.} When a single step cannot be resolved by the locator, the executor invokes one LLM call to handle that step, then resumes mechanical replay for subsequent steps. This avoids the ${\sim}60$\,s cost of a full Layer~1 re-execution. Twelve rounds used this mechanism, all completing successfully with a mean of 5.0 LLM calls, significantly fewer than the 11.6 calls required by full Layer~1 execution.

Together, these two mechanisms rescued 26 rounds (17.3\% of the experiment) from full fallback, each saving an estimated 60\,s of LLM execution time.

\subsection{Checker-in-the-Loop}

A subtle advantage of Layer~2 replay is its immunity to LLM false-completion signals. In the baseline, the LLM decides when a task is ``Done'', and this judgment is unreliable: 8 of 57 baseline failures (14.0\%) occur when the LLM declares completion but the ground-truth checker rejects the result (e.g., the LLM tapped a search suggestion instead of executing the query, or navigated to a settings screen without toggling the switch). SkillDroid experiences \textbf{zero} such false-completion failures: Layer~2 replay completes when all skeleton steps have executed (a mechanical condition, not an LLM judgment), and the checker verifies the final state. Even in Layer~1 rounds, the checker catches premature completion and continues execution. This design eliminates an entire failure category that accounts for 14\% of baseline failures.

\end{document}